# Simulation and Performance Analysis of Adaptive Filtering Algorithms in Noise Cancellation

**Lilatul Ferdouse1, Nasrin Akhter2, Tamanna Haque Nipa3 and Fariha Tasmin Jaigirdar4**

**1 Department of Computer Science, Stamford University Bangladesh
Dhaka, Bangladesh**

**2 Department of Computer Science, Stamford University Bangladesh
Dhaka, Bangladesh**

**3 Department of Computer Science, Stamford University Bangladesh
Dhaka, Bangladesh**

**4 Department of Computer Science, Stamford University Bangladesh
Dhaka, Bangladesh**

**Abstract**
Noise problems in signals have gained huge attention due to the need of noise-free output signal in numerous communication systems. The principal of adaptive noise cancellation is to acquire an estimation of the unwanted interfering signal and subtract it from the corrupted signal. Noise cancellation operation is controlled adaptively with the target of achieving improved signal to noise ratio. This paper concentrates upon the analysis of adaptive noise canceller using Recursive Least Square (RLS), Fast Transversal Recursive Least Square (FTRLS) and Gradient Adaptive Lattice (GAL) algorithms. The performance analysis of the algorithms is done based on convergence behavior, convergence time, correlation coefficients and signal to noise ratio. After comparing all the simulated results we observed that GAL performs the best in noise cancellation in terms of Correlation Coefficient, SNR and Convergence Time. RLS, FTRLS and GAL were never evaluated and compared before on their performance in noise cancellation in terms of the criteria we considered here.
***Keywords:*** *Adaptive Filter, Noise, Mean Square Error, RLS, FTRLS, GAL, Convergence*

## 1. Introduction

A Digital communication system consists of a transmitter, channel and receiver connected together. Typically the channel suffers from two major kinds of impairments: Intersymbol interference and Noise. The principle of noise cancellation is to obtain an estimate of the interfering signal and subtract it from the corrupted signal. Adaptive noise cancellation [1]-[2], a specific type of interference cancellation, relies on the use of noise cancellation by subtracting noise from a received signal, an operation controlled in an adaptive manner for the purpose of improved signal to noise ratio. It is basically a dual-input, closed loop adaptive control system as illustrated in fig 1.

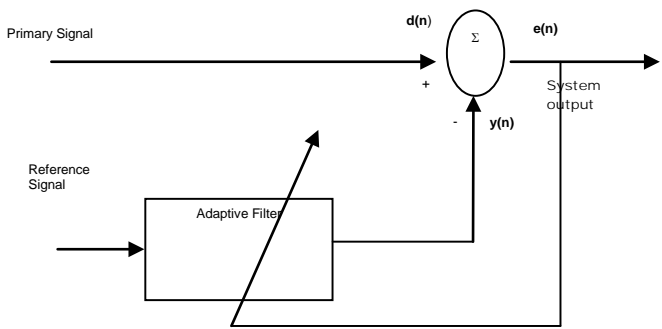

Fig. 1  Noise Cancellations.

Here the adaptive filter [2] is used to cancel unknown interference contained in a primary signal, with the cancellation being optimized in some sense. The primary signal serves as the desired response for the adaptive filter. The reference signal is employed as the input to the filter. This paper studies and analyzes the performances of three adaptive algorithms in noise cancellation. We present simulations based on different types of signals mixed with various types of noise. Despite the theoretical nature of the study, efforts have been made to emphasize signals of practical use. Therefore, audio and electrical signals that are subject to noise in real world have been considered. Audio files were read and microphones connected real audio signals. The analysis of the results offered useful insight into the characteristics of the algorithms

## 2. Literature Review

Different applications of the adaptive digital filters were studied as early as the 1950s. Now adaptive filters are ubiquitous tools for numerous real-world scientific and industrial applications. The initial works on adaptive echo cancellers started around 1965. It appears that Kelly of Bell Telephone Laboratories was the first to propose the use of an adaptive filter for echo cancellation, with the speech signal itself utilized in performing the adaptation [2]-[3]. Yuu-Seng Lau, Zahir M.Hossain and Richard Harrisa explained  Performance of Adaptive Filtering Algorithms: A comparative study. It showed a cooperative performance study between the time-varying LMS (TV-LMS) and other two main adaptive approaches: The Least Mean Square (LMS) algorithm and the Recursive Least Square (RLS) algorithm. Their study disclosed the algorithm execution time, the minimum Mean Square Error (MSE) and required filter order [4]. Hyun-Chool Shin,Ali H.Sayed and Woo-Jin Song described the Mean Square Performance of Adaptive filters using averaging theory. This paper uses averaging analysis to study the mean-square performance of adaptive filters, not only in terms of stability conditions but also in terms of expressions for the mean-square error and the mean-square deviation of the filter, as well as in terms of the transient performance of the corresponding partially averaged systems [5]. R.C. North J.R. Zeidler , T.R. Albert , W.H. Ku presented  the Comparison of adaptive lattice filters to LMS transversal filters for sinusoidal cancellation. This paper compare the performance of the recursive least squares lattice (RLSL) and the normalized step-size stochastic gradient lattice (SGL) algorithms to that of the least mean square (LMS) transversal algorithm for the cancellation of sinusoidal interference. It is found that adaptive lattice filters possess a number of advantages over the LMS transversal filter, making them the preferred adaptive noise cancellation (ANC) filter structure if their increased computational costs can be tolerated [6]. Syed .A.Haider and M.Lotfizad developed a new approach for canceling attenuating noise in speech signals.  This paper presented a nice tradeoff between convergence properties and computational complexity and showed that the convergence property of fast affine projection (FAP) adaptive filtering algorithm is superior to that of usual LMS, NLMS, and RLS algorithm [7].

For the learning of FIR filters using linear adaptive filtering algorithms ,it is well known that recursive-least-squares(RLS) algorithms produce a faster convergence speed than stochastic gradient descent techniques, such as the basic least-mean-squares(LMS) algorithms, or even gradient-adaptive-lattice LMS(GAL)[2]-[3].In this paper we present an implementation of adaptive noise canceller using Recursive Least Square (RLS), Fast Transversal Recursive Least Square (FTRLS) and Gradient Adaptive Lattice (GAL) algorithms with the intention to compare their performance in noise cancellation in terms of convergence behavior, convergence time, and correlation coefficients and signal to noise ratio.

## 3. Adaptive Algorithms

Recursive Least Squares (RLS) algorithm is capable of realizing a rate of convergence that is much faster than the LMS algorithm, because the RLS algorithm utilizes all the information contained in the input data from the start of the adaptation up to the present.

### 3.1 The Standard RLS Algorithms

In the method of least squares, at any time instant $n > 0$ the adaptive filter parameter (tap weights) are calculate so that the quantity of the cost function

$$\varsigma(n) = \sum_{k=1}^{n} \rho_n(k) e_n^2(k) \quad (1)$$

is minimized and hence the name least squares. In k=1 is the time at which the algorithm starts, $e_n(k), k = 1,2,...n$, are the samples of error estimates that would be obtained if the filter were run from time $k = 1$ to n ,using the set of filter parameters that is computed at time n, and $\rho_n(k)$ is a weighting function. Actually the RLS algorithm performs the following operations:

- Filters the input signal $x(n)$ through the adaptive filter $w(n-1)$ to produce the filter output $y(n)$
- Calculates the error sample $e(n) = d(n) - y(n)$
- Recursively updates the gain vector $k(n)$
- Updates the adaptive filter coefficients

3.1.1 The algorithm can be summarized through following steps:

Input Parameters
Tap-weight vector estimate, $\hat{w}(n-1)$
Input vector, $x(n)$
Desired output, $d(n)$
And the matrix, $\psi_\lambda^{-1}(n-1)$
Output:
Filter output, $y_{n-1}(n-1)$
Tap weight vector update, $\hat{w}(n)$
And the updated matrix, $\psi_\lambda^{-1}(n)$
Procedure:
1. Computation of the gain vector:

$$u(n) = \psi_\lambda^{-1}(n-1)x(n)$$

$$k(n) = \frac{1}{\lambda + x^T(n)u(n)} u(n)$$

2. Filtering:

$$\hat{y}_{n-1}(n) = \hat{w}^T(n-1)X(n)$$

3. Error-estimation:

$$\hat{e}_{n-1}(n) = d(n) - \hat{y}_{n-1}(n)$$

4. Tap-weight vector adaptation:

$$\hat{w}(n) = \hat{w}(n-1) + k(n)\hat{e}_{n-1}(n)$$

5. $\psi_\lambda^{-1}(n)$ Update:

$$\psi_\lambda^{-1}(n) = \lambda^{-1}(\psi_\lambda^{-1}(n-1) - k(n)[x^T(n)\psi_\lambda^{-1}(n-1)])$$

### 3.2 Fast Transversal RLS Algorithm

Fast transversal filter (FTF) algorithm involves the combined use of four transversal filters for forward and backward predictions, gain vector computation and joint process estimation. The main advantage of FTF algorithm is reduced computational complexity as compared to other available solutions. The derivation of the algorithm follows hereafter.

### 3.2.1 Summary of the FTRLS Algorithm

The FTRLS algorithm is summarized below by collecting together the relevant equations.

Input parameters:
Tap-input vector $x_{N-1}(n-1)$, desired output $d(n)$
Tap-weight vectors $\bar{a}_N(n-1), \bar{g}_N(n-1)$ and $\hat{w}_N(n-1)$
Normalized gain vector, $\bar{k}_N(n-1)$
Least squares sums or auto correlations $\varsigma_N^{ff}(n-1), \varsigma_N^{bb}(n-1)$

Output:
The updated values of
$\bar{a}_N(n), \bar{g}_N(n), \hat{w}_N(n), \bar{k}_N(n), \varsigma_N^{ff}(n), \varsigma_N^{bb}(n)$

Prediction:
$$f_{N,n-1}(n) = \hat{a}_N^T(n)x_{N+1}(n)$$
$$f_{N,n}(n) = \gamma_N(n-1)f_{N,n-1}(n)$$
$$f_{N,n}(n) = \gamma_N(n-1)f_{N,n-1}(n)$$
$$\varsigma_N^{ff}(n) = \lambda\varsigma_N^{ff}(n-1)f_{N,n-1}(n)$$

$$\gamma_{N+1}(n) = \lambda\frac{\varsigma_N^{ff}(n-1)}{\varsigma_N^{ff}(n)}\gamma_N(n-1)$$

$$\overline{K}_{N+1}(n) = \begin{bmatrix}0\\ \overline{K}_N(n-1)\end{bmatrix} + \lambda^{-1}\frac{f_{N,n-1}(n)}{\varsigma_N^{ff}(n)}\tilde{a}_N(n-1)$$

$$\tilde{a}_N(n) = \tilde{a}_N(n-1) - \begin{bmatrix}0\\ \overline{K}_N(n-1)\end{bmatrix}f_{N,n}(n)$$

$$b_{N,n-1}(n) = \lambda\varsigma_N^{bb}(n-1)\overline{K}_{N+1,N-1}(n)$$

$$\beta(n) = 1 - b_{N,n-1}(n)\gamma_{N+1}(n)\overline{K}_{N+1,n-1}(n)$$
$$\gamma_N(n) = \beta^{-1}(n)\gamma_{N-1}(n)$$
$$b_{N,n}(n) = \gamma_N(n)b_{N,n-1}(n)$$
$$\varsigma_N^{bb}(n) = \lambda\varsigma_N^{bb}(n-1) + b_{N,n}(n)b_{N,n-1}(n)$$
$$\begin{bmatrix}\mathrm{K}_N(n)\\ 0\end{bmatrix} = \overline{\mathrm{K}}_{N+1}(n) - \overline{\mathrm{K}}_{N+1,N-1}(n)\overline{g}_N(n-1)$$

$$\bar{g}_N(n) = \bar{g}_N(n-1) - \begin{bmatrix}\bar{k}_N(n)\\ 0\end{bmatrix}b_{N,n}(n)$$

Filtering:
$$e_{N,n-1}(n) = d(n) - \hat{w}_N^T(n-1)x_N(n)$$
$$e_{N,n}(n) = \gamma_N(n)e_{N,n-1}(n)$$
$$\hat{w}_N(n) = \hat{w}_N(n-1) + \bar{k}_N(n)e_{N,n}(n)$$

### 3.3 Gradient Adaptive Lattice

The *gradient-adaptive lattice(GAL) filter* is due to Griffiths(1977,1978) and may be viewed as a natural extension of the normalized least-mean-square(LMS) filter in that both types of filter rely on a stochastic gradient approach for their algorithmic implementations.

### 3.3.1 Summary of the GAL Algorithm

Parameters: $M$=final prediction order
$\quad\quad\quad\beta$ =constant, lying in the range (0.1)
$\quad\quad\quad\hat{\mu} < 0.1$
$\quad\quad\quad\delta$ : small positive constant
$\quad\quad\quad a$ : another small positive constant

Multistage lattice predictor:
$$f_0(m) = b_0(n) = u(n)$$
$$\varepsilon_{m-1}(n) = \beta\varepsilon_{m-1}(n-1) + (1-\beta)(|f_{m-1}(n)|^2 + |b_{m-1}(n-1)|^2)$$
$$f_m(n) = f_{m-1}(n) + k_m b_{m-1}(n-1)$$
$$b_m(n) = b_{m-1}(n-1) + k_m f_{m-1}(n)$$
$$\hat{k}_m(n) = \hat{k}_m(n-1) - \frac{\hat{\mu}}{\varepsilon_{m-1}(n)}\left(f_{m-1}^*(n)b_m(n) + b_{m-1}(n-1)f_m^*(n)\right)$$

Filtering:
$$y_m(n) = y_{m-1}(n) + \hat{h}_m^*(n)b_m(n)$$
$$e_m(n) = d(n) - y_m(n)$$
$$\|b_m(n)\|^2 = \|b_{m-1}(n)\|^2 + |b_m(n)|^2$$
$$\hat{h}_m(n+1) = \hat{h}_m(n) + \frac{\tilde{\mu}}{\|b_m(n)^2\|^2}b_m(n)e_m^*(n)$$

## 4. Simulation Results

Simulation based on four different types of signals mixed with various types of noise. Signals are periodic signal, audio signal, chirp signal and saw-tooth signal. Each signal has been subjected to some noise. Then the convergence behaviors of the RLS, FTF and GAL algorithms for these signals have been analyzed. Audio files were read and microphones connected real audio signals. The signals were then polluted by white, pink, grey and burst noise. We also apply AWGN channel model and take low, moderate and high signal-to-noise ratio. The signals were then passed through the simulation of the adaptive filter, and their error recovery rate, correlation coefficient and time were calculated. The analysis of the results offered useful insight into the characteristics of the algorithms. For the RLS algorithm, two parameters were varied to find their effect on the performance. One of them is the filter length, and the other is the forgetting factor. In Fast RLS algorithms, FTF and GAL, the performances were analyzed by varying different filter length, forgetting factor and step size parameter.

### 4.1 Comparison based on Noise Cancellation Performance

In order to compare noise cancellation capability, three methods of presentation have been shown. One of them is the plotting of Mean Square Error with number of samples. Error convergence characteristics of the three algorithms have been shown on the same graph to attain visual comparison.

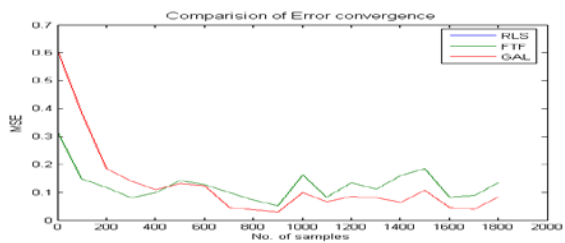

Fig. 2  Comparison of noise cancellation for RLS. FTF, GAL algorithms for sinusoidal signal.

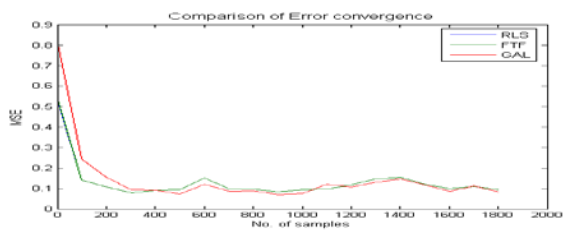

Fig. 3  Comparison of noise cancellation for RLS. FTF, GAL algorithms for saw-tooth signal.

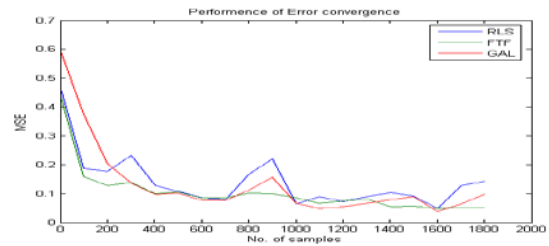

Fig. 4  Comparison of noise cancellation for RLS. FTF, GAL algorithms for Chirp signal.

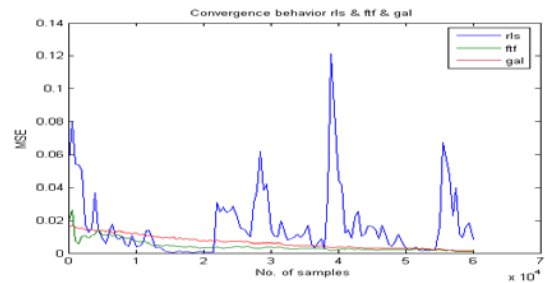

Fig. 5  Comparison of noise cancellation for RLS. FTF, GAL algorithms for audio signal.

In the case of periodic signal (mixed with white noise), RLS and FTF algorithms perform better than GAL and they almost show same convergence behavior. But, GAL's performance is not satisfactory in this case. The same thing is repeated also for the sawtooth signal when it is corrupted by white noise and the chirp signal which is distorted by pink noise. Concluding by audio noise, which is corrupted by white noise it can be told that FTF performs the best, then comes GAL and after that RLS comes.

### 4.2 Comparison of Correlation Coefficient

A tabular method is used to compare the correlation coefficients of the algorithms. In this comparison, the algorithms have been compared for the same signal to noise ratio combination

Table 1: Comparison of correlation coefficients of RLS, FTF, GAL for different signals mixed with white noise

| Correlation Coefficients | | | |
|---|---|---|---|
| *Signal Types* | *RLS* | *FTF* | *GAL* |
| *Chirp* | 0.8401 | 0.8501 | 0.9218 |
| *Sinusoidal* | 0.9464 | 0.9459 | 0.9422 |
| *Saw tooth* | 0.8935 | 0.8909 | 0.9021 |
| *audio* | 0.9798 | 0.9988 | 0.9989 |

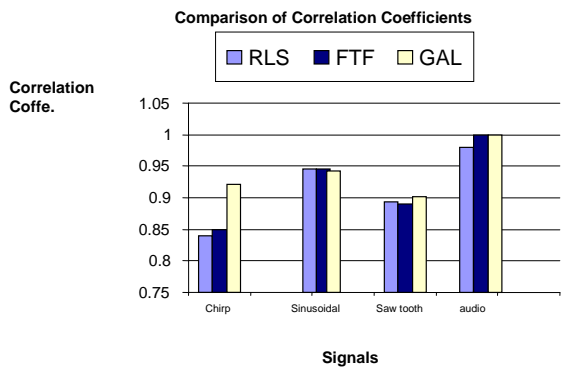

Fig. 6 Comparison of correlation coefficients of RLS. FTF, GAL for different signals mixed with white noise.

The above chart (Fig. 6) reveals the decision that Gradient adaptive lattice (GAL) has the best noise cancellation performance since the correlation coefficient for all signals except sinusoidal signal than other two algorithms. This means that GAL has achieved close approximation of the desired signal in all cases. On other side RLS and FTF show same performance in case of sinusoidal signal

4.3 Comparison of Convergence Time

Taking the same signal to noise ratio, the performance of the RLS, FTF and GAL algorithms are compared in terms of their convergence time, which is given in tabular and graphical form. Analyzing fig. 7, it is revealed that RLS takes the least convergence time than the other two. GAL takes the second position in this occurrence. Lastly comes the FTF algorithm that requires more convergence time than standard RLS and GAL algorithms

Table 2: Comparison of convergence time of RLS, FTF, GAL for different signals mixed with white noise

| Convergence Time (S) | | | |
|---|---|---|---|
| *Signal Type* | RLS | FTF | GAL |
| *Chirp signal* | 0.453 | 1.797 | 0.672 |
| *Sinusoidal signal* | 0.641 | 2.438 | 0.844 |
| *Saw tooth signal* | 0.422 | 1.453 | 1.453 |

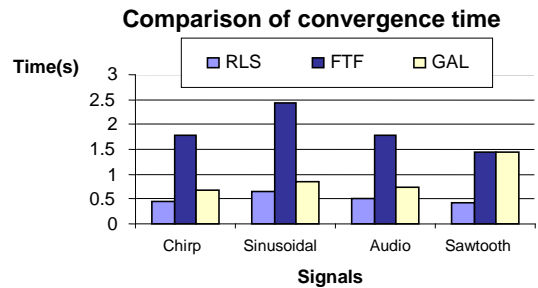

Fig. 7 Comparison of convergence time of RLS. FTF, GAL for different signals mixed with white noise.

4.4 Comparison of Signal-to-Noise Ratio.

The experiment is divided into three parts: In part 1, the input signals-to-noise ratio (SNR) is high; in part 2, it is mid and in part 3, it is low.

Part 1:

Table 3: Comparison of signal to noise ratio of RLS, FTF, GAL algorithms when given SNR=30dB (high)

| Signal to noise ratio (30dB) | | | |
|---|---|---|---|
| *Signal Types* | RLS | FTF | GAL |
| *Chirp* | 13.9296 | 24.7287 | 68.74 |
| *Sinusoidal* | 14.5297 | 14.513 | 72.7454 |
| *Saw tooth* | 12.7979 | 12.788 | 71.474 |
| *Audio* | 13.0794 | 25.513 | 46.6549 |

Part 2:

Table 4: Comparison of signal to noise ratio of RLS, FTF, GAL algorithms when given SNR=10dB (mid)

*Signal to noise ratio (10dB)*

| Signal Types | RLS | FTF | GAL |
|---|---|---|---|
| Chirp | 9.7591 | 8.355 | 19.3497 |
| Sinusoidal | 7.7829 | 7.7703 | 18.1702 |
| Saw tooth | 7.5157 | 7.5087 | 20.4481 |
| Audio | 8.2704 | 9.3365 | 10.0652 |

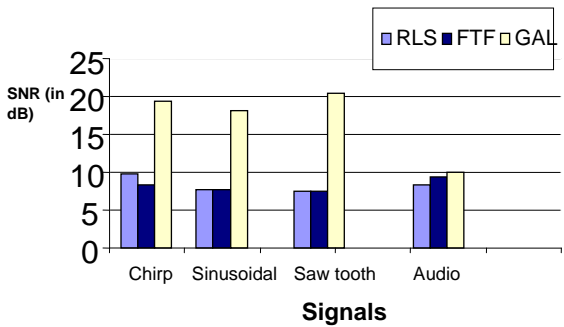

Fig. 8 Comparison of signal to noise ration of RLS. FTF, GAL algorithms when SNR=10dB (mid).

Part 3:

Table 5: Comparison of signal to noise ratio of RLS, FTF, GAL algorithms when given SNR=-10dB (low)

*Signal-to-noise ratio(-10db)*

| Signal Types | RLS | FTF | GAL |
|---|---|---|---|
| Chirp | -9.2383 | -9.3945 | -8.7799 |
| Sinusoidal | -9.7834 | -9.7822 | -9.4333 |
| Saw tooth | -9.2232 | -9.2027 | -9.0513 |
| Audio | -9.8538 | -10.0622 | -9.3343 |

Based On the simulated results, the following facts have been noted.

- For a fixed signal-to-noise ratio of 30 dB and considering different type of signals, we analyze the noise cancellation performance of the three algorithms. In all cases, Gradient Adaptive Algorithms (GAL) shows the best performance. In case of chirp signal, periodic signal and sawtooth signal, the output SNR of the GAL algorithms is higher than double of the input SNR value that is approximately 70dB. For audio signal, it is about 50dB.
- Almost in every case of noise cancellation, RLS and FTF algorithm show worse performance than GAL algorithms. For both of the algorithms, the output SNR values are decreased.

The same occurrence happens for both 10dB and –10dB respectively in the case of mid and low SNR. The results presented in Table 3 and 4 clearly show the superior value of the output SNR of the GAL over other two algorithms.

## 5. Conclusions

The components of adaptive noise canceller were generated by computer simulation using MATLAB. The analysis of the results offered useful insight into the characteristics of the algorithms. For the RLS algorithm, two parameters were varied to find their effect on the performance. One of them is the filter length, and the other is the forgetting factor. In Fast RLS algorithms, FTRLS and GAL, the performances were analyzed by varying different filter length, forgetting factor and step size parameter. The study revealed that, for the RLS, FTRLS, GAL algorithms, the increase in filter length results in increased MSE and increased convergence time. For step size, such generalization cannot be made. If the step size is increased, the algorithms converge faster. But the error tends to become unstable. Forgetting factor is other parameter which also controls the stability and the rate of convergence. Typically, it has been seen that a forgetting factor, ranges between .99 to 1, gives satisfactory results. When it comes to convergence time, the length of the filter is a big factor. It takes the RLS and FTRLS significantly longer time to compute of coefficient increases. To draw a comparison among three algorithms, the main factors that should be kept in mind are noise cancellation performance, convergence time and making the signal to noise ratio high. It is found in all cases that RLS has performed as medium level in canceling noise. Fast RLS algorithms have achieved more effective noise cancellation. In some cases FTRLS may have taken slightly more time to converge, but its error has always dipped down below that of the RLS algorithms. In the case of convergence time, GAL algorithm shows the best performance among three algorithms. The situations in which the amplitude or frequency in signal encounters abrupt changes, the RLS and FTRLS algorithms show poor performance. In these cases, RLS and FTRLS graphs show sudden rise of error whereas the GAL remains stable to zero. Signal-to-noise ratio can be increased by canceling noise from signal and providing more strength to signal. In this case GAL always shows better performance and enhances SNR value in all types of signal either the SNR is high, low or mid. In the end, it can be stated that Fast RLS algorithms especially GAL should be preferred over the standard RLS for noise cancellation unless error convergence time, output SNR is a matter of great concern.

**Lilatul Ferdouse** studied Computer Science and Engineering at University of Dhaka, Bangladesh from 1999 to 2003. In 2003 she received the Bachelor degree. She received the M. S. degree in Computer Science and Engineering in 2004 from University of Dhaka. She is working as a Senior Lecturer in Department of Computer Science, Stamford University Bangladesh. Her area of interest includes Digital Signal Processing, Data Mining, and Wireless Communication.

**Nasrin Akhter** studied Computer Science and Engineering at University of Dhaka, Bangladesh from 1999 to 2003. In 2003 she received the Bachelor degree. She received the M. S. degree in Computer Science and Engineering in 2004 from University of Dhaka. She is working as an Assistant Professor in Department of Computer Science, Stamford University Bangladesh. Her area of interest includes Digital Signal Processing, Data Mining, Cryptography and Security.

**Tamanna Haque Nipa** studied Computer Science and Engineering at Stamford University Bangladesh 1999 to 2003. In 2003 she received the Bachelor degree. Currently she is studying M.S. at Bangladesh University of Engineering and Technology. She is working as a Lecturer in Department of Computer Science, Stamford University Bangladesh. Her area of interest includes Digital Signal Processing and Wireless Communication.

**Fariha Tasmin Jaigirdar** studied Computer Science and Engineering at Chittagong University of Engineering and Technology, Bangladesh from 2001 to 2005. In 2005 she received the Bachelor degree. Currently she is studying M.S. at Bangladesh University of Engineering and Technology. She is working as a Lecturer in Department of Computer Science, Stamford University Bangladesh. Her area of interest includes Digital Signal Processing and Wireless Networking.